\documentclass[12pt]{article}
\usepackage{amsbsy,graphicx}

\def\ga{\gamma}
\def\de{\delta}
\def\ep{\epsilon}

\def\kap{\kappa}

\def\th{\theta}

\def\vp{\varphi}

\def\Ga{\Gamma}
\def\De{\Delta}

\def\half{{1\over{2}}}
\def\pd{\partial}
\def\del{{\bf\nabla}}

\newcommand{\beq}{\begin{equation}}
\newcommand{\eeq}{\end{equation}}
\newcommand{\bea}{\begin{eqnarray*}}
\newcommand{\eea}{\end{eqnarray*}}
\newcommand{\beaq}{\begin{eqnarray}}
\newcommand{\eeaq}{\end{eqnarray}}
\newcommand{\vph}{\boldsymbol{\varphi}}
\newcommand{\vrh}{\boldsymbol{\rho}}
\newcommand{\val}{\boldsymbol{\alpha}}

\newcommand{\vaa}{\boldsymbol{a}}
\newcommand{\va}{{\bf a}}
\newcommand{\vP}{{\bf P}}
\newcommand{\vQ}{{\bf Q}}
\newcommand{\bfe}{{\bf e}}
\newcommand{\bfv}{{\bf v}}
\newcommand{\bds}[1]{\boldsymbol{#1}}
\newcommand{\G}{\mathcal{G}}
\newcommand{\T}{\mathcal{T}}
\begin{document}
\begin{flushright}APCTP-1999012\\KIAS-P99055\end{flushright}
\centerline{\Large \bf Reflection Amplitudes of ADE Toda Theories}
\centerline{\Large \bf and Thermodynamic Bethe Ansatz}
\vskip 1cm
\centerline{\large Changrim Ahn$^{1}$, V. A. Fateev$^{2}$,
Chanju Kim$^{3}$, Chaiho Rim$^{4}$, and Bedl Yang$^{1}$}
\vskip 1cm
\centerline{\it$^{1}$Department of Physics, Ewha Womans University}
\centerline{\it Seoul 120-750, Korea}
\vskip .5cm
\centerline{\it$^{2}$Laboratoire de Physique Math\'ematique,
Universit\'e Montpellier II}
\centerline{\it Place E. Bataillon, 34095 Montpellier, France}
\centerline{\it and}
\centerline{\it Landau Institute for Theoretical Physics}
\centerline{\it Kossygina 2, 117334 Moscow, Russia}
\vskip .5cm
\centerline{\it $^{3}$ School of Physics, Korea Institute for Advanced Study}
\centerline{\it Seoul, 130-012, Korea}
\vskip .5cm
\centerline{\it $^{4}$ Department of Physics, Chonbuk National University}
\centerline{\it Chonju 561-756, Korea}
\vskip 1cm
\centerline{\small PACS: 11.25.Hf, 11.55.Ds}
\vskip 2cm
\centerline{\bf Abstract}
We study the ultraviolet asymptotics in affine Toda theories.
These models are considered as perturbed non-affine Toda theories.
We calculate the reflection amplitudes, which relate different
exponential fields with the same quantum numbers.
Using these amplitudes we derive the quantization condition for the 
vacuum wave function, describing zero-mode dynamics, and
calculate the UV asymptotics of the effective central charge.
These asymptotics are in a good agreement with thermodynamic
Bethe ansatz results.

\newpage
\section{Introduction}

There is a large class of 2D quantum field theories (QFTs) which
can be considered as perturbed conformal field theories (CFTs) \cite{sasha}.
These theories are completely defined if one specifies its CFT data
and the relevant operator which plays the role of perturbation.
The CFT data contain explicit information about ultraviolet (UV)
asymptotics of the field theory while its long distance
property is the subject of analysis.
If a perturbed CFT contains only massive particles, it is 
equivalent to the relativistic scattering theory and is completely
defined by specifying the $S$-matrix.
Contrary to CFT data the $S$-matrix data exibit some information
about long distance properties of the theory in an explicit way,
while the UV asymptotics have to be derived.

A link between these two kinds of data would provide a good view point
for understanding the general structure of 2D QFTs.
In general this problem does not look tractable.
Whereas the CFT data can be specified in a relatively simple way,
the general $S$-matrix is very complicated object even in 2D.
However, there exists an important class of 2D QFTs (integrable theories)
where scattering theory is factorized and $S$-matrix can be described in
great details.

In this case one can apply the nonperturbative methods based on the 
$S$-matrix data.
One of these methods is thermodynamic Bethe ansatz (TBA) \cite{yang,alyosha}.
It gives the possibility to calculate the ground state energy $E(R)$
(or effective central charge $c_{\rm eff}(R)$) for the system on the
circle of size $R$.
At small $R$ the UV asymptotics of $c_{\rm eff}(R)$ can be compared 
with that following from the CFT data.

Usually the UV asymptotics for the effective central charge can be 
derived from the conformal perturbation theory.
In this case the corrections to $c_{\rm eff}(0)=c_{\rm CFT}$ have a form of 
series in $R^{\gamma}$ where $\gamma$ is defined by the dimension
of perturbing operator.
However, there is an important class of QFTs where the UV asymptotics
of $c_{\rm eff}(R)$ is mainly determined by the zero-mode dynamics
(see for example \cite{roaming, FOZ, ZamZam, AKR}).
In this case the UV corrections to $c_{\rm CFT}$ have the form of series
in inverse powers of $\log(1/R)$.
This UV expansion is also encoded in CFT data \cite{ZamZam}.

The simplest integrable QFT with the logarithmic expansion for the 
effective central charge is the sinh-Gordon (ShG) model, which
is an integrable deformation of Liouville conformal field theory (LFT).
It was shown in paper \cite{ZamZam} that the crucial role in the
description of the zero-mode dynamics in the ShG model is played by
the ``reflection amplitude'' of the LFT, which determine
the asymptotics of the ground state wave function in this theory.
(The reflection amplitudes in CFT define the linear transformations
between different exponential fields, corresponding to the same 
primary field of chiral algebra.)
The perturbative term in the ShG model restricts the dynamics to
the interval of size $l\sim\log(1/R)$, which leads to the quantization
condition for the momentum $P$ conjugated to the zero-mode.
The solution $P(R)$ of the quantization condition determines all 
logarithmic terms in the UV asymptotics of the effective central charge
$c_{\rm eff}(R)$.
The similar approach was used in \cite{AKR} to describe the UV asymptotics
of Bullough-Dodd and supersymmetric ShG models.
In all cases it was found perfect agreement with TBA results based on
the $S$-matrix data.

In this paper we study the UV behaviour of the effective central charge
in affine Toda field theories (ATFTs) associated with simply-laced
Lie algebras $G=ADE$.
The number of particles in the ATFT is equal to the rank $r$ OF $G$.
For $r>>1$ the numerical analysis of the TBA equations, especially
in the UV region, becomes very complicated.
The analytical consideration of the TBA equations permits to calculate
only the first term (${\cal O}(1/l^2)$) in the UV expansion for 
$c_{\rm eff}(R)$ which also contains undetermined constants 
\cite{martins,FKS}.
So it is useful to have the full logarithmic expansion for this function
with explicitly determined coefficients from the CFT data.

In the next section we describe the scattering theory and give 
the exact relations between the parameters of the action and
spectrum of the ATFT.
This QFT can be considered as perturbed CFT, namely, perturbed
non-affine Toda theory (NATT), which possesses nontrivial chiral
algebra ($W(G)$-algebra).
We derive the reflection amplitudes for this CFT.
In the weak coupling limit these amplitudes reduce to the coefficients
in the asymptotics of the wave function of the one-dimensional 
open quantum Toda chain (see for example \cite{OlsPer}).
This system describes the ``semiclassical'' limit of the zero-mode
dynamics of the NATT.
The relation between the reflection amplitudes and the vacuum wave
function of the NATT is studied in sect.3.
In sect.4 we derive the quantization condition and calculate the 
UV asymptotics for the effective central charge for the ATFT.
In sect.5 we compare this asymptotics with numerical data.
An independent derivation of quantization condition for one-dimensional
quantum Toda chain is given in Appendix.

\section{Affine and Non-Affine Toda Theories, Normalization Factors 
and Reflection Amplitudes}

The ATFTs corresponding to Lie algebra  $G$ is 
described by the action
\beq
{\cal A}=\int d^2x\left[{1\over{8\pi}}(\pd_{\mu}\vph)^2
+\mu\sum_{i=1}^{r}e^{b\bfe_i\cdot\vph}+\mu e^{b\bfe_0\cdot\vph}
\right],\label{action}
\eeq
where $\bfe_i,\ i=1,\ldots,r$ are the simple roots of the Lie algebra
$G$ of rank $r$ and $-\bfe_0$ is a maximal root satisfying
\beq
\bfe_0+\sum_{i=1}^{r}n_i\bfe_i=0.
\label{maxroot}
\eeq
The fields in Eq.(\ref{action}) are normalized so that at $\mu=0$
\beq
\langle\varphi_a(x)\varphi_b(y)\rangle=-\delta_{ab}\log|x-y|^2.
\label{twopoint}
\eeq
We will consider later the case of simply-laced Lie algebras
$G=A, D, E$. 

For real $b$ the spectrum of these ATFTs consists of $r$ particles 
with the masses $m_i$ ($i=1,\ldots,r$) given by
\beq
m_i={\overline m}\nu_i
\eeq
where 
\beq
{\overline m}^2={1\over{2h}}\sum_{i=1}^{r}m_i^2
\eeq
and here $h$ is Coxeter number and $\nu_i^2$ are the eigenvalues of 
the mass matrix:
\beq
M_{ab}=\sum_{i=1}^{r}n_i(\bfe_i)^a(\bfe_i)^b+(\bfe_0)^a(\bfe_0)^b.
\eeq

The relation between the parameter ${\overline m}$ characterizing 
the spectrum of physical particles and parameter $\mu$ in the action
(\ref{action}) can be obtained by Bethe ansatz method
(see for example \cite{mumass,fateev}).
It can be easily derived from the results of \cite{fateev} and has
the form:
\beq
-\pi\mu\ga(1+b^2)=\left[{{\overline m}k(G)\Ga\left({1\over{
(1+b^2)h}}\right)\Ga\left(1+{b^2\over{(1+b^2)h}}\right)\over{
2\Ga(1/h)}}\right]^{2(1+b^2)}
\label{mmu}
\eeq
where $\ga(x)=\Ga(x)/\Ga(1-x)$ and
\beq
k(G)=\left(\prod_{i=1}^{r}n_i^{n_i}\right)^{1/2h}
\label{kofg}
\eeq
with $n_i$ defined in Eq.(\ref{maxroot}).

For the practical applications it is useful to have the relations
between the parameter ${\overline m}$ and the minimal masses $m_1$ in $A,E$ 
and masses $m_n=m_{n-1}$ in $D_n$ ATFTs:
\beaq
&A_{n-1}: m_1^2=4\sin^{2}(\pi/n){\overline m}^{2},\qquad
&D_n:m_n^{2}=m_{n-1}^{2}=m^2=2{\overline m}^{2},\label{dnmass}\\
&E_6:m_1^{2}=(3-\sqrt{3}){\overline m}^{2},\qquad
&E_7:m_1^{2}=8\sin^{2}(\pi/9){\overline m}^{2},\nonumber\\
&E_8:m_1^2=8\sqrt{3}\sin(\pi/5)\sin(\pi/30){\overline m}^2.\nonumber
\eeaq

The scattering matrix for the particles $m_i$ in ATFT was constructed
in \cite{AFZ,BCD}. It depends on one parameter
\beq
B={b^2\over{1+b^2}}
\label{coupling}
\eeq
and is invariant under duality transformation $b\to 1/b$ or
$B\to 1-B$.
This scattering matrix is a pure phase (See, for example, \cite{Oota,FKS}):
\beaq
S_{ij}(\th)&=&\exp(-i\delta_{ij}(\th))\qquad{\rm where}
\label{Satft}\\
\delta_{ij}(\th)&=&\int_{0}^{\infty}
{dt\over{t}}\left[8\sinh\left({\pi Bt\over{h}}\right)
\sinh\left({\pi(1-B)t\over{h}}\right)
{\left(2\cosh{\pi t\over{h}}-{\bf I}\right)^{-1}}_{ij}
-2\de_{ij}\right]\sin(\th t),\nonumber
\eeaq
where ${\bf I}$ is the incident matrix defined by
${\bf I}_{ij}=2\de_{ij}-\bfe_i\cdot\bfe_j$.

The ATFTs can be considered as perturbed CFTs. 
Without the last term with the zeroth root $\bfe_0$, the action
in Eq.(\ref{action}) describes the NATT,
which is conformal.
To describe the generator of conformal symmetry we introduce
the complex coordinates $z=x_1+i x_2$ and ${\overline z}=x_1-i x_2$
and vector
\beq
\vQ=Q\vrh,\qquad Q=b+1/b, \qquad \vrh=\half\sum_{\val>0}\val \eeq
where the sum in the definition of the Weyl vector $\vrh$ runs over
all positive roots $\val$ of $G$.

The holomorphic stress-energy tensor
\beq
T(z)=-\half(\pd_z\vph)^2+\vQ\cdot\pd_z^2\vph
\eeq
ensures the local conformal invariance of the NATT with the central
charge
\beq
c=r+12\vQ^2=r\left(1+h(h+1)Q^2\right).
\label{central}
\eeq

Besides the conformal invariance the NATT possesses extended
symmetry generated by $W(G)$-algebra.
The full chiral $W(G)$-algebra contains $r$ holomorphic fields
$W_j(z)$ ($W_2(z)=T(z)$) with spins $j$ which follows the
exponents of Lie algebra $G$. 
The explicit representation of these fields in terms of fields
$\pd_z\vph$ can be found in \cite{FL}.
The primary fields $\Phi_w$ of $W(G)$ algebra are classified
by  $r$ eigenvalues $w_j,\ j=1,\ldots,r$ of the operator
$W_{j,0}$ (the zeroth Fourier component of the current  $W_j(z)$):
\beq
W_{j,0}\Phi_w=w_j\Phi_w,\qquad
W_{j,n}\Phi_w=0,\quad n>0.
\eeq
The exponential fields
\beq
V_{\vaa}(x)=e^{(\vQ+\vaa)\cdot\vph(x)}
\label{vertex}
\eeq
are spinless conformal primary fields with dimensions
\beq
\Delta(\vaa)=w_2(\vaa)={\vQ^2\over{2}}-{\vaa^2\over{2}}.
\eeq
The fields in Eq.(\ref{vertex}) are also primary fields 
with respect to all chiral algebra $W(G)$ with the eigenvalues
$w_j$ depending on $\vaa$. 
The functions $w_j(\vaa)$ which define the representation of
$W(G)$-algebra possess the symmetry with respect to the Weyl
group ${\cal W}$ of Lie algebra $G$ \cite{FL}, i.e.
\beq
w_j({\hat s}\vaa)=w_j(\vaa);\qquad
{\hat s}\in {\cal W}.
\eeq
It means that the fields $V_{{\hat s}\vaa}$ for different
${\hat s}\in {\cal W}$ are reflection images of each other
and are related by the linear transformation:
\beq
V_{\vaa}(x)=R_{\hat s}(\vaa)V_{{\hat s}\vaa}(x)
\label{reflection}
\eeq
where $R_{\hat s}(\vaa)$ is the ``reflection amplitude''.

This function is an important object in CFT and plays a 
crucial role in the calculation of one-point functions
in perturbed CFT \cite{FLZZ}.
To calculate the function $R_{\hat s}(\vaa)$, we introduce
the fields $\Phi_w$:
\beq
\Phi_w(x)=N^{-1}(\vaa)V_{\vaa}(x)
\label{phiw}
\eeq
where normalization factor $N(\vaa)$ is chosen in the way that
field $\Phi_w$ satisfies the conformal normalization condition
\beq
\langle\Phi_w(x)\Phi_w(y)\rangle={1\over{|x-y|^{4\De}}}.
\label{normal}
\eeq
The normalized fields $\Phi_w$ are invariant under reflection
transformations and hence;
\beq
R_{\hat s}(\vaa)={N(\vaa)\over{N({\hat s}\vaa)}}.
\eeq
For the calculation of the normalization factor $N(\vaa)$, we
can use the integral representation for the correlation functions
of the $W(G)$-invariant CFT. (See \cite{FL} for details.) 
We note that operators ${\widehat Q}_i$ defined as
\beq
{\widehat Q}_i=\mu\int d^2x e^{b\bfe_i\cdot\vph(x)}
\eeq
commute with all of the elements of $W(G)$-algebra and can be
used as screening operators for the calculation of the correlation
functions in the NATT.
If parameters $\vaa$ satisfy the condition
\beq
2\vQ+2\vaa+\sum_{i=1}^{r}k_i\bfe_i=0
\eeq
with non-negative integer $k_i$, we obtain form Eqs.(\ref{phiw}) and 
(\ref{normal})
the following expression for the function $N(\vaa)$ in terms of
Coulomb integrals \cite{FL}:
\beq
N^2(\vaa)=|x|^{4\De}\left\langle
V_{\vaa}(x)V_{\vaa}(0)\prod_{i=1}^{r}{{\widehat Q}_i^{k_i}\over{
k_i!}}\right\rangle
\label{coulomb}
\eeq
where the expectation value in Eq.(\ref{coulomb}) is taken
over the Fock vacuum of massless fields $\vph$ with the
correlation functions (\ref{twopoint}).

The normalization integral can be calculated and the result has
the form:
\beq
N^2(\vaa)=\left(\pi\mu\ga(b^2)\right)^{-2\vrh\cdot(\vQ+\vaa)/b}
\prod_{\val>0}{\Ga(1+Q_{\val}/b)\Ga(1+Q_{\val}b)
\Ga(1+a_{\val}/b)\Ga(1+a_{\val}b)\over{
\Ga(1-Q_{\val}/b)\Ga(1-Q_{\val}b)\Ga(1-a_{\val}/b)\Ga(1-a_{\val}b)}}
\label{naresult}
\eeq
in terms of the scalar products
\beq \label{qalpha}
Q_{\val}=\vQ\cdot\val,\qquad a_{\val}=\vaa\cdot\val
\eeq
where the product runs over all positive roots of Lie algebra $G$.

We accept Eq.(\ref{naresult}) as the proper analytical continuation
of the function $N^2(\vaa)$ for all $\vaa$. 
It gives us the following expression for the reflection
amplitude $R_{\hat s}(\vaa)$:
\beq
R_{\hat s}(\vaa)={N(\vaa)\over{N({\hat s}\vaa)}}=
{A_{{\hat s}\vaa}\over{A_{\vaa}}}
\label{rhats}
\eeq
where 
\beq
A_{\vaa}=\left(\pi\mu\ga(b^2)\right)^{\vrh\cdot\vaa/b}
\prod_{\val>0}\Ga(1-a_{\val}/b)\Ga(1-a_{\val}b).
\eeq
The reflection relation Eq.(\ref{reflection}) can be written
in more symmetric form as:
\beq \label{reflection2}
A_{\vaa}V_{\vaa}(x)=A_{{\hat s}\vaa}V_{{\hat s}\vaa}(x),\qquad
{\hat s}\in{\cal W}.
\eeq
In following we will be interested in the values of functions
$A_{\vaa}$ for imaginary $\vaa=i\vP$.
We denote as $V(\vP,x)=V_{i\vP}(x)$ and
\beq
A(\vP)=A_{i\vP}=\left(\pi\mu\ga(b^2)\right)^{i\vP\cdot\vrh/b}
\prod_{\val>0}\Ga(1-iP_{\val}/b)\Ga(1-iP_{\val}b).
\label{ampliti}
\eeq
Using these objects we can construct the combination which is
invariant under the Weyl reflections:
\beq
\Psi_{\vP}=\sum_{{\hat s}\in {\cal W}}A({\hat s}\vP)
V({\hat s}\vP).
\label{primary}
\eeq

\section{Reflections of Quantum Mechanical Waves}

In this section we follow the LFT analysis \cite{ZamZam} to interpret 
the relation between the primary fields of the NATTs and the wave 
functionals $\Psi[\vph(x)]$
whose asymptotic behaviours are described by the wave functions of
the zero-modes.
The zero-modes of the fields $\vph(x)$ are defined as:
\beq
\vph_0 = \int_0^{2\pi}\vph(x) \frac{dx_1}{2\pi}.
\eeq
Here we consider the NATT on an infinite plane cylinder of circumference
$2\pi$ with coordinate $x_2$ along the cylinder playing the role
of imaginary time.
In the asymptotic region where the potential terms in the NATT action 
become negligible (${\bf e}_{i}\cdot\vph_0\to-\infty$ for all $i$), 
the fields can be expanded in terms of free field operators $\va_n$
\beq
\vph(x)=\vph_0-{\cal P}(z-{\overline z})+
\sum_{n\neq 0}\left({i\va_n\over{n}}e^{inz}+
{i{\overline\va}_n\over{n}}e^{-in{\overline z}}\right),
\label{free}
\eeq
where ${\cal P}=-i\del_{\vph_0}$ is the conjugate momentum of $\vph_0$.
In this region any state of the NATT can be decomposed into a direct 
product of two parts, namely, a wave function of the zero-modes and a state 
in Fock space generated by the operators $\va_n$.
In particular, the wave functional corresponding to the primary state
Eq.(\ref{primary}) can be expressed as a direct product of a wave function 
of the zero-modes $\vph_0$ and Fock vacuum: 
\beq
\Psi_{\vP}[\vph(x)]\sim \Psi_{\vP}(\vph_0)\otimes\vert 0\rangle
\label{vip}
\eeq
where the wave function $\Psi_{\vP}(\vph_0)$ in this asymptotic region
is a superposition of plane waves with momenta ${\hat s}\vP$.

The reflection amplitudes of the NATT defined in the previous section
can be interpreted as those for the wave function of the zero-modes
in the presence of potential walls.
This can be understood most clearly in the semiclassical limit $b\to 0$
where one can neglect the operators $\va_n$ in Eq.(\ref{free})
even for significant values of the parameter $\mu$.
The full quantum effect can be implemented simply by introducing
the exact reflection amplitudes which take into account also non-zero-mode
contributions \cite{ZamZam}.
The resulting Schr\"odinger equation is given by
\beq
\left[{r\over{12}}-\del_{\vph_0}^2
+\mu\sum_{i=1}^{r}e^{b\bfe_i\cdot\vph_0}\right]
\Psi_{\vP}(\vph_0) = E_0\Psi_{\vP}(\vph_0)
\eeq
with the ground state energy
\beq
E_0=-{r\over{12}}+\vP^2.
\label{energy}
\eeq
Here the momentum $\vP$ is any continuous real vector.
The effective central charge can be obtained from Eq.(\ref{energy})
where $\vP^2$ takes the minimal possible value for the perturbed theory.
Since only asymptotic form of the wave function matters, we derive
the reflection amplitudes of the ATFTs in the way that we need 
only the LFT result.

In the $\mu \rightarrow 0$ limit which will be of our interest,
the potential vanishes almost everywhere 
except for the values of $\vph_0$ where some of exponential terms in the 
potential become large enough to overcome the small value of $\mu$. 
In this case, each exponential term
$e^{b \bfe_i\cdot\vph_0}$ in the interaction represent a wall 
with $\bfe_i$ being its normal vector. 
If we consider the behaviour of a wave function near a wall normal to
$\bfe_i$ where the effect of other interaction terms becomes negligible, 
the problem becomes equivalent to the LFT in the $\bfe_i$ direction.
The potential becomes flat in the $(r-1)$-dimensional orthogonal 
directions.
The asymptotic form of the energy eigenfunction is then given by the 
product of that of Liouville wave function and $(r-1)$-dimensional plane wave,
\beaq \label{nearwall}
\Psi &\sim& \left[ e^{i P_i\vp_{0i}} + S_L(P_i) e^{-i P_i\vp_{0i}} \right]
e^{i\vP_\perp\cdot\vph_0} \nonumber\\
&\sim& e^{i \vP\cdot\vph_0} + S_L(P_i)e^{i \hat{s}_i\vP\cdot\vph_0}\,,
\eeaq
where $\hat{s}_i$ denotes the Weyl reflection by the simple root $\bfe_i$
and $P_i$ the component of $\vP$ along $\bfe_i$ direction.
$S_L(P)$ is the reflection amplitude of the LFT,
\beq \label{liouville}
S_L(P)=\left(\pi\mu\gamma(b^2)\right)^{-iP/b}{\Gamma\left(1+iPb\right)
\Gamma\left(1+iP/b\right)\over{\Gamma\left(1-iPb\right)
\Gamma\left(1-iP/b\right)}}.
\eeq
Since the wave function interpretation makes sense only in the semiclassical
limit, it is the $b\to 0$ limit of Eq.(\ref{liouville}) which 
can be obtained from the solution of the Schr\"odinger equation for the LFT. 

We can see from Eq.(\ref{nearwall}) that the momentum of the
reflected wave by the $i$-th wall is given by the Weyl reflection 
$\hat{s}_i$ acting on the incoming momentum. 
If we consider the reflections from all the potential walls, 
the wave function in the asymptotic region is a superposition of the
plane waves reflected by potential walls in different ways.
The momenta of these waves form the orbit of the Weyl group 
${\cal W}$ of the Lie algebra $G$; 
\beq \label{wavefunction}
\Psi_\vP(\vph_0)\simeq\sum_{\hat{s}\in {\cal W}}
A(\hat{s}\vP)e^{i\hat{s}\vP\cdot\vph_0}.
\eeq
This is indeed the wave function representation of the primary field 
(\ref{primary}) in the asymptotic region.

It follows from Eq.(\ref{nearwall}) that the amplitudes $A(\vP)$ satisfy 
the relations
\beq \label{ratio}
\frac{A(\hat{s}_i\vP)}{A(\vP)} = S_L(P_i).
\eeq
For a general Weyl element ${\hat s}$ which can be represented by a product 
of the Weyl elements ${\hat s}_i$ associated with the simple roots
by $\hat{s} = \hat{s}_{i_k}\hat{s}_{i_{k-1}}\cdots\hat{s}_{i_1}$,
the above equation can be generalized to
\beq
\frac{A(\hat{s}_{i_k}\cdots\hat{s}_{i_1}\vP)}{A(\vP)}
= S_L(\vP\cdot\bfe_{i_1})S_L(\hat{s}_{i_1}\vP\cdot\bfe_{i_2})
S_L(\hat{s}_{i_{2}}\hat{s}_{i_1}\vP\cdot\bfe_{i_3})
\cdots S_L(\hat{s}_{i_{k-1}}\cdots\hat{s}_{i_1}\vP\cdot\bfe_{i_k}).
\label{genref}
\eeq
Using the properties of the Weyl group (see for example \cite{humphreys}) 
and the explicit form of the amplitude $S_{L}(P)$,
it is straightforward to verify that the following function $A(\vP)$ 
satisfies Eqs.(\ref{ratio}) and (\ref{genref}):
\beq \label{amplitude}
A(\vP) = \left(\pi\mu\gamma(b^2)\right)^{i\vrh\cdot\vP/b}
\prod_{\val>0}\Gamma\left(1-iP_{\val}b\right)
\Gamma\left(1-iP_{\val}/b\right)
\eeq
where $P_{\val}= \val\cdot\vP$ is a scalar product with a positive root 
$\val$.  
The fact that Eq.(\ref{amplitude}) is similar to (\ref{ampliti})  illustrates 
the relation between the primary fields and zero modes wave functions in 
NATT.

\section{Quantization Condition and Scaling Function for ATFT}

The analysis of the previous section can be used to obtain 
the scaling functions in the deep UV region of the ATFTs defined 
on a cylinder with circumference $R\rightarrow 0$. 
The additional potential term in the ATFT Lagrangian corresponding to 
the zeroth root $\bfe_0$ introduces new potential wall in that direction
(see Fig.1 as a simplest example, the $A_2$ ATFT).
With this addition, the region of $\vph_0$ made of the non-affine Toda 
potential walls (Weyl chamber) is now closed and the momentum of the wave 
function should be quantized depending on the size of the enclosed region.
This quantized momentum defines the scaling function
$c_{\rm eff}$ in the UV region by Eq.(\ref{energy}).

\begin{figure}[t]
\includegraphics[height=9cm]{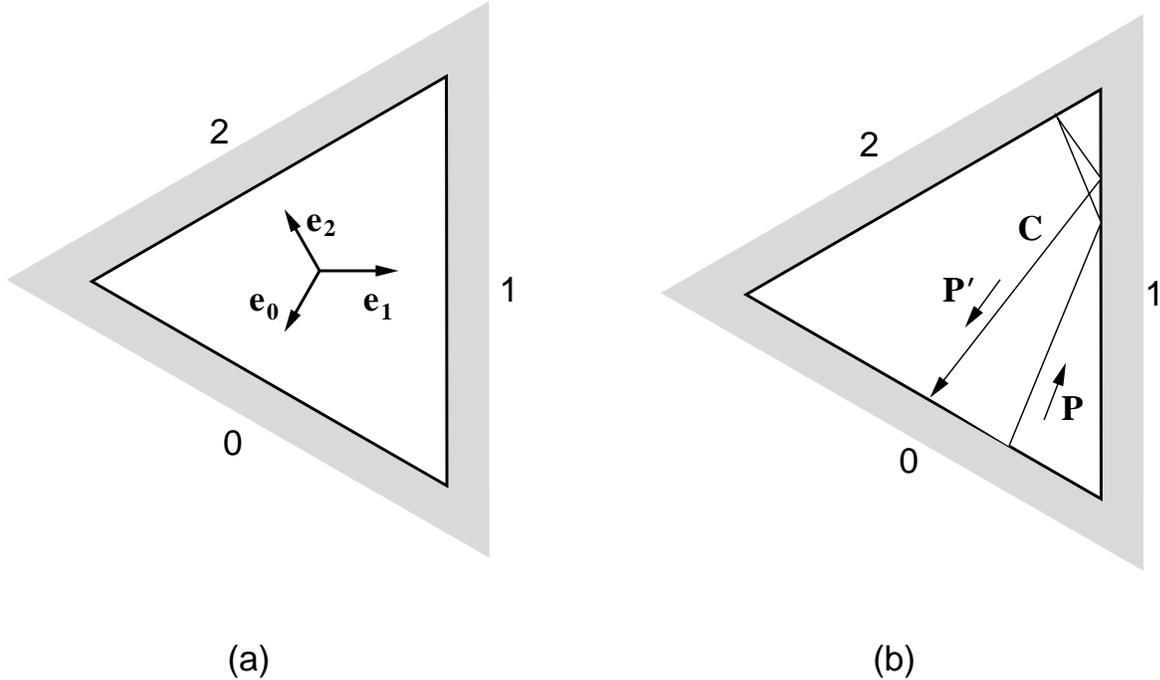}
\caption{\small (a) Potential walls in $A_2$ affine Toda theory.
(b) A wave with momentum $\vP$ near the zeroth wall comes back 
to the same wall with momentum $\vP'$ after a series of reflections.}
\vskip 1cm
\end{figure}

The quantization condition can be derived using the arguments of sect.3.  
For the moment we assume that the circumference of the cylinder is $2\pi$. 
Consider the path $C$ of a wave which starts with momentum $\vP$ 
and comes back (after a series of reflections by other walls) to the zeroth 
potential wall with momentum $\vP'$.
It will then be reflected by the zeroth wall.
Fig.1(b) illustrates a multiple reflection in the two-dimensional potential. 
To satisfy the self-consistency condition, 
the momentum $\vP'$ after the last reflection by the
zeroth wall should be equal to the incoming momentum $\vP$
so that $\hat{s}_0\vP'=\vP$. 
Furthermore, since the zeroth wall is again Liouville-type, 
the momenta $\vP' = \hat{s}_0\vP$ of the {\em incident} wave 
and $\vP$ of the {\em reflected} wave should satisfy Eq.(\ref{ratio}) 
which leads to
\beq \label{ratiozero}
\frac{A(\hat{s}_0\vP)}{A(\vP)} = S_L(\vP\cdot\bfe_0).
\eeq
On the other hand, since $\hat{s}_0$ is given by a product of 
the Weyl reflections corresponding to simple roots, each representing   
the reflection experienced by the wave along the path $C$, 
the left hand side of Eq.(\ref{ratiozero}) can be obtained from 
Eq.(\ref{amplitude}). 
Therefore, Eq.(\ref{ratiozero}) gives a nontrivial quantization condition 
for the momentum $\vP$.
This condition can be generalized using the same arguments for other
potential walls instead of the zeroth one. Then we obtain
\beq \label{ratios}
\frac{A(\hat{s}_0\hat{s}\vP)}{A(\hat{s}\vP)} = S_L(\hat{s}\vP\cdot\bfe_0)\,.
\eeq
where $\hat{s}$ is an arbitrary Weyl group element.

Using Eq.(\ref{amplitude}) we can write (\ref{ratios}) 
\beq \label{quant1}
\left(\pi\mu\gamma(b^2)\right)^{i\vP\cdot\hat{s}\bfv/b}
\left[\prod_{\val>0}\frac{G(\hat{s}\vP\cdot\hat{s}_0\val)}
{G(\hat{s}\vP\cdot\val)}\right]
\frac{G(\hat{s}\vP\cdot\bfe_0)}{G(-\hat{s}\vP\cdot\bfe_0)} = 1\,,
\eeq
where
\[
\bfv = \hat{s}_0\vrh -\vrh+\bfe_0=-(\bfe_0\cdot\vrh)\bfe_0 + \bfe_0 
= h \bfe_0
\]
and we define a function
\[
G(P)=\Gamma\left(1-iP/b\right)\Gamma\left(1-iPb\right).
\]
The $\Gamma$-function factors in Eq.(\ref{quant1}) can be further simplified.
First, consider the action of $\hat{s}_0$ on a positive root
$\val$: $\hat{s}_0\val = \val - \bfe_0 (\bfe_0\cdot\val)$, which is
either $\val$ or $\val+\bfe_0$ if $\val\neq -\bfe_0$
since $-\bfe_0$ is the maximal root.
In the first case, the factor $G(\hat{s}\vP\cdot\hat{s}_0\val)$ in
Eq.(\ref{quant1}) is cancelled out by the same factor in the denominator,
while, in the second case, there is no cancellation since $\val+\bfe_0$ is
a negative root. Finally, $\hat{s}_0\bfe_0 = -\bfe_0$ and
%\marginpar{$\Longleftarrow$}%
the corresponding factor $G(\hat{s}\vP\cdot\bfe_0)$ appears twice in
Eq.(\ref{ratios}). Using the property $\bfe_0\cdot\val = 0$ or $1$ for
$\val > 0$ ($\val \neq - \bfe_0$) and $\bfe_0 \cdot \bfe_0 = 2$, we can
simplify Eq.(\ref{quant1}) as
\beq \label{quant2}
\left(\pi\mu\gamma(b^2)\right)^{ih\vP\cdot\hat{s}\bfe_0/b}
\prod_{\val>0}\left[\frac{G(-\vP\cdot\hat{s}\val)}{G(\vP\cdot\hat{s}\val)}
\right]^{-\hat{s}\val\cdot\hat{s}\bfe_0} = 1\,.
\eeq

Since the Weyl element $\hat{s}$ is arbitrary, Eq.(\ref{quant2}) 
leads to the following condition for the lowest energy state 
\beq \label{quantization}
2hQL\vP = 2\pi\vrh - \sum_{\val>0}\val\de(P_{\val})\,,
\eeq
where
\[
L = -\frac{1}{2(1+b^2)}\log[\pi\mu\gamma(b^2)]\,,
\]
and
\beq \label{delta}
\de(P) = -i \log\frac{\Gamma(1+i P/b)\Gamma(1+i P b)}%
{\Gamma(1-i P/b)\Gamma(1-i Pb)}\,.
\eeq
This is the quantization condition for the momentum $\vP$ in the
$\mu\rightarrow 0$ limit. We see that effectively each positive root $\val$
causes a phase shift of Liouville type.

Now we consider the system defined on a cylinder with the circumference $R$.
When we scale back the size from $R$ to $2\pi$,
the parameter $\mu$ in the action (\ref{action}) changes to
\beq	
\mu \rightarrow \mu \left(\frac{R}{2\pi}\right)^{2(1+b^2)}.
\eeq
The $\mu\rightarrow 0$ limit is realized as the deep UV limit $R\to 0$.
The rescaling changes the definition of $L$ in Eq.(\ref{quantization}) by
\beq \label{L}
L=-\log\frac{R}{2\pi}-\frac{1}{2(1+b^2)}\log[\pi\mu\gamma(b^2)]\,.
\eeq
The ground state energy with the circumference $R$ is given by
\beq
E(R)=-{\pi c_{\rm eff}\over{6R}}\quad
{\rm with}\quad
c_{\rm eff}=r-12\vP^2
\label{grenergy}
\eeq
where $\vP$ satisfies Eq.(\ref{quantization}).

In this limit, Eq.(\ref{quantization}) can be solved perturbatively. 
For this we expand the function $\de(P)$ in Eq.(\ref{delta}) 
in powers of $P$,
\beq \label{expand}
\de(P) = \de_1(b)P + \de_3(b)P^3 + \de_5(b) P^5 \cdots\,,
\eeq
where
\beaq \label{deltas}
\de_1(b) &=& -2Q\gamma_E \,,\nonumber\\
\de_3(b) &=& \frac23\zeta(3)(b^3 + b^{-3}) \,, \nonumber\\
\de_5(b) &=& -\frac25\zeta(5)(b^5 + b^{-5}).
\eeaq
Using the relation 
\[
\sum_{\val>0} (\val)^a(\val)^b = h\de^{ab},
\] 
we obtain
\[
hl\vP = 2\pi\vrh-\de_3(b)\sum_{\val>0}\val(P_{\val})^3
-\de_5(b)\sum_{\val>0}\val(P_{\val})^5 - \cdots\,,
\]
with
\beq \label{smalll}
l\equiv 2QL + \de_1\,.
\eeq
The above equation can be solved iteratively in powers of $1/l$. 
Inserting the solution into Eq.(\ref{grenergy}), we find
\beq
c_{\rm eff}=r-\frac{r(h+1)}{h}\left[\left(\frac{2\pi}{l}\right)^2
-\frac{24\delta _3}{2\pi}\left(\frac{2\pi}{l}\right)^5D_4(G)
-\frac{24\delta _5}{2\pi}\left(\frac{2\pi}{l}\right)^7D_6(G)
+{\cal O}(l^{-8})
\right]
\label{ceff}
\eeq
where the coefficients $D(G)$ are given in terms of a scalar product 
$\rho_{\val}=\vrh\cdot\val$ by
\[
D_4(G) = \frac{1}{r(h+1)h^4}\sum_{\val >0}\rho_{\val}^4\,, \qquad
D_6(G) = \frac{1}{r(h+1)h^6}\sum_{\val >0}\rho_{\val}^6\,.
\]
The values of $D_4(G)$ and $D_6(G)$ can be evaluated explicitly:
\beaq \label{d4d6}
&&D_4(A_{n-1})={(2n^2-3)\over 60n^2}\,,\qquad
  D_6(A_{n-1})={(n^2-2)(3n^2-5)\over 168n^4}\,,\nonumber\\
&&D_4(D_{n})={(16n^3-45n^2+27n+8)\over 480(n-1)^3}\,,\nonumber\\
&&D_6(D_{n})
   ={(48n^5-213n^4+262n^3+6n^2-101n-32)\over 84(2(n-1))^5}\,,\nonumber\\
&&D_4(E_n)={r(h+1)\over 24(r+2)h}\,, \qquad (n=6,\ 7,\ 8)\,,\nonumber\\
&&D_6(E_6)={43\cdot 73\over 8\cdot (12)^4}\,,\nonumber\\
&&D_6(E_7)={514001\over 14\cdot (18)^5}\,,\nonumber\\
&&D_6(E_8)={(31)^2\over 54\cdot(30)^2}\,.
\label{explicit}
\eeaq
Using the relation (\ref{mmu}) between the parameter $\mu$ in the action
and the parameter ${\overline m}$ characterizing the spectrum, 
we can represent the effective central 
charge (\ref{ceff}) for the simply-laced ATFTs in terms of
dimensionless value ${\overline m}R$.
In this form it can be compared with TBA results.

We note that expansion (\ref{ceff},\ref{explicit}) for $D_n$ 
ATFTs can be continued to the half-integer $n$.
The $D_n$ Toda theories with half-integer $n=l+1/2$ were
studied in \cite{DGZ}.
Besides the bosonic fields $\vph$ these models include the Majorana
fermion $\psi$.
The corresponding scattering theory is also self-dual ($b\to 1/b$).
To obtain the values of $c_{\rm eff}(R)$ for these models we should
express $c_{\rm eff}(R)$ in terms of parameter $mR$ where
$m$ is defined by Eq.(\ref{dnmass}) and take the continuation
for $r=n$, $h=2(n-1)$, and 
$k(D_n)=2^{{n-3\over{2(n-1)}}}$ to the half-integer $n=l+1/2$.
In particular, for $n=3/2$ we obtain exactly the expansion for
the supersymmetric ShG model found in \cite{AKR}.

\vskip 1cm
\leftline{\underline{\bf Example: $A_2$ ATFT}}
\vskip .5cm
As a simplest example, we consider the $A_2$ ATFT more explicitly. 
In this case, there are two simple roots
$\bfe_1$ and $\bfe_2$, and the affine root $\bfe_0$ is given by
$\bfe_0 = -(\bfe_1 + \bfe_2)$. For each simple root, a Liouville-type
potential wall is placed in the normal direction and the walls form
a regular triangle as shown in Fig.1.

As the coefficient $\mu\to 0$, the size of the triangle becomes 
large and the zero-mode dynamics reduces to the
quantum mechanical problem in two dimensions surrounded by a triangular 
potential wall
where the potential vanishes except the vicinity of the walls. 
The energy eigenstate of the system is then the standing wave in the triangle 
which is a superposition of plane waves with momenta reflected by walls 
as explained in sect.3. 
Since the Weyl group ${\cal W}$ of $A_2$ algebra consists from six elements,
${\cal W}= \{1,\hat{s}_1,\hat{s}_2,\hat{s}_1\hat{s}_2,\hat{s}_2\hat{s}_1,
\hat{s}_1\hat{s}_2\hat{s}_1=\hat{s}_2\hat{s}_1\hat{s}_2=\hat{s}_0\}$,
the wave function can be written as a sum of six plane waves generated by
the Weyl reflections as in Eq.(\ref{wavefunction}) with their coefficients
determined by Liouville reflection amplitudes. 
For example, if one considers
the path $C$ of Fig.1 followed by a wave with momentum $\vP$, one finds
\bea
A(\hat{s}_1\vP) &=& S_L(\vP\cdot \bfe_1) A(\vP)\,,\\
A(\hat{s}_2\hat{s}_1\vP) &=& S_L(\hat{s}_1\vP\cdot \bfe_2) 
A(\hat{s}_1\vP)
= S_L(\hat{s}_1\vP\cdot \bfe_2)S_L(\vP\cdot \bfe_1) A(\vP)\,, \\
A(\hat{s}_1\hat{s}_2\hat{s}_1\vP)
&=& S_L(\hat{s}_2\hat{s}_1\vP\cdot \bfe_1) A(\hat{s}_2\hat{s}_1\vP)
= S_L(\hat{s}_2\hat{s}_1\vP\cdot \bfe_1)
S_L(\hat{s}_1\vP\cdot \bfe_2) S_L(\vP\cdot \bfe_1) A(\vP)\,,
\eea
which can be summarized as Eq.(\ref{amplitude}).

If the wave is further reflected by the zeroth wall, we have
\beaq \label{a2quant}
A(\hat{s}_0\hat{s}_1\hat{s}_2\hat{s}_1\vP)
&=& S_L(\hat{s}_1\hat{s}_2\hat{s}_1\vP\cdot \bfe_0)
A(\hat{s}_1\hat{s}_2\hat{s}_1\vP) \nonumber \\
&=& S_L(\hat{s}_1\hat{s}_2\hat{s}_1\vP\cdot \bfe_0)
S_L(\hat{s}_2\hat{s}_1\vP\cdot \bfe_1)
S_L(\hat{s}_1\vP\cdot \bfe_2) S_L(\vP\cdot \bfe_1) A(\vP) \,.
\eeaq
Since $\hat{s}_0=\hat{s}_1\hat{s}_2\hat{s}_1$, Eq.(\ref{a2quant}) gives a
quantization condition for $\vP$ which is the special case of
Eq.(\ref{quant1}). 
Simplifying the arguments of $S_L$ in Eq.(\ref{a2quant}),
we find
\[
S_L(\vP \cdot \bfe_1) S_L(\vP \cdot \bfe_2)
S_L(\vP \cdot (\bfe_1+\bfe_2))^2 = 1\,,
\]
which is equivalent to Eq.(\ref{quant2}).

\section{Comparison with the TBA results}

A standard approach to study the scaling behaviour of integrable QFTs is
to solve the TBA equations. 
In this section we compute the scaling functions of the ATFTs 
in the UV region from the TBA equations and compare
them with the results in Eq.(\ref{ceff}) based on the reflection 
amplitudes.

The TBA equations for the ATFTs are given by ($i=1, \cdots, r$)
\beq
m_i R \cosh\theta=\ep_i (\theta,R)+
\sum_{j=1}^r\int\varphi_{ij} (\theta-\theta') 
\log\left(1+ e^{-\ep_i(\th')}\right){d\theta'\over 2\pi},
\label{toda_tba}
\eeq
where $\varphi_{ij}$ is the kernel which is equal to the logarithmic 
derivative of the $S$-matrix $S_{ij} (\th)$ in Eq.(\ref{Satft})
\[
\varphi_{ij} (\th)=-i {d \over d\th}\log S_{ij} (\th)=\de'_{ij}(\th).
\]
These are $r$-coupled nonlinear integral equations for 
the `pseudo-energies' $\ep_i(\theta,R)$ which give 
the scaling function of the effective central charge 
\beq  \label{ceff_tba}
c_{\rm eff}(R) = \sum_{i=1}^r\frac{3Rm_i}{\pi^2}
\int \cosh\theta \log\left(1+ e^{-\ep_i(\th)}\right)d\theta.
\eeq

It is quite difficult task to solve the TBA equations analytically and
compare directly with Eq.(\ref{ceff}).
To obtain higher order terms in $1/l$ expansion, one needs to solve
complicated coupled nonlinear differential equations.
Even the lowest order terms at the order of $1/l^2$ contain constants
which can not be decided by the scattering data.
While the method used in \cite{AKR} may be applicable here, 
we will concentrate only on numerical analysis of the TBA equations
to avoid any digression to different problem.

Even the numerical analysis is limited if the rank $r$ grows
since a large number of equations amplify the numerical errors
entering in the iteration procedure.
Therefore we will consider a few ATFTs with low ranks in each series of
A-D-E, namely, $A_2$, $A_3$, $A_4$, $D_4$ and $E_6$ ATFTs.
The effective central charge $c_{\rm eff}(R)$ is computed 
by solving Eq.(\ref{ceff_tba}) iteratively as a function of ${\overline m}R$. 
In order to compare the numerical data with our results based on
the reflection amplitudes, 
we fit the numerical data for $c_{\rm eff}(R)$ from the TBA equations 
for many different values of $R$ with the function (\ref{ceff}) where
$\de_1$, $\de_3$ and $\de_5$ are considered as the fitting parameters. 
For this comparison the relation
(\ref{mmu}) between the parameter $\mu$ in the action and parameter
${\overline m}$ for the particle masses is used.
These parameters $\de_i$'s are then compared with Eq.(\ref{deltas}) 
defined from the reflection amplitude of the LFT. 
Since we already separate out the dependence on the Lie algebra $G$,
our numerical results for the parameters $\de_i$'s should be 
independent of $G$.

Tables 1--3 show the values of parameters $\de_i$'s obtained numerically 
from TBA equations for different values of the coupling constant $B$ 
in $A_2$, $A_3$, $A_4$, $D_4$ and $E_6$ ATFTs. 
We see that they are in excellent agreement with those values of $\de_i$'s 
following from the reflection amplitudes supplemented with Eq.(\ref{mmu}).
Thus numerical TBA analysis fully supports the validity of our whole scheme
based on the reflection amplitude, $\mu$-${\overline m}$ relation and 
the quantization condition on $\vP$. 

\begin{table}
\centerline{
\begin{tabular}{||c||c||c|c|c|c|c||} \hline
\rule[-.4cm]{0cm}{1.cm}B
& $\de^{\rm (RA)}_1$ & $\de_1^{\rm (TBA)}(A_2)$
& $\de_1^{\rm (TBA)}(A_3)$ & $\de_1^{\rm (TBA)}(A_4)$
& $\de_1^{\rm (TBA)}(D_4)$ & $\de_1^{\rm (TBA)}(E_6)$ \\ \hline
0.20 & --2.88608 & --2.88608 & --2.884 &  &  &
       \\
0.25 & --2.66604 & --2.66604 & --2.66603 & --2.664   &   &
        \\
0.30 & --2.51918 & --2.51918 & --2.51918 & --2.51915 & --2.5186 &
        \\
0.35 & --2.42035 & --2.42035 & --2.42035 & --2.42035 & --2.42033 &
         \\
0.40 & --2.35647 & --2.35647 & --2.35647 & --2.35647 & --2.35647 &
       --2.3568 \\
0.45 & --2.32049 & --2.32049 & --2.32049 & --2.32049 & --2.32049  &
       --2.3208 \\
0.50 & --2.30886 & --2.30886 & --2.30886 & --2.30886 & --2.30886 &
      --2.30886 \\ \hline
\end{tabular}
}
\caption{$\de_1^{\rm (RA)}$ vs.
$\de_1^{\rm (TBA)}$ for $A_2,A_3, A_4, D_4$, and $E_6$ ATFTs.}
\end{table}
\begin{table}
\centerline{
\begin{tabular}{||c||c||c|c|c|c|c||} \hline
\rule[-.4cm]{0cm}{1.cm}B
& $\de^{\rm (RA)}_3$ & $\de_3^{\rm (TBA)}(A_2)$
& $\de_3^{\rm (TBA)}(A_3)$ & $\de_3^{\rm (TBA)}(A_4)$
& $\de_3^{\rm (TBA)}(D_4)$ & $\de_3^{\rm (TBA)}(E_6)$ \\ \hline
0.20 & 6.51114 & 6.509   &         &  &  &
       \\
0.25 & 4.31827 & 4.3180  & 4.310   &         &   &
               \\
0.30 & 3.08111 & 3.08100 & 3.08113 & 3.06    & 2.92    &
        \\
0.35 & 2.34480 & 2.34473 & 2.34484 & 2.3445  & 2.3438  &
         \\
0.40 & 1.90842 & 1.90839 & 1.90845 & 1.90845 & 1.90759 &
       1.93  \\
0.45 & 1.67590 & 1.67588 & 1.67592 & 1.67592 & 1.67589 &
       1.69  \\
0.50 & 1.60274 & 1.60273 & 1.60276 & 1.60276 & 1.60276 &
       1.6028 \\ \hline
\end{tabular}
}
\caption{$\de_3^{\rm (RA)}$ vs.
$\de_3^{\rm (TBA)}$ for $A_2,A_3, A_4, D_4$, and $E_6$ ATFTs.}
\end{table}
\begin{table}
\centerline{
\begin{tabular}{||c||c||c|c|c|c|c||} \hline
\rule[-.4cm]{0cm}{1.cm}B
& $\de^{\rm (RA)}_5$ & $\de_5^{\rm (TBA)}(A_2)$
& $\de_5^{\rm (TBA)}(A_3)$ & $\de_5^{\rm (TBA)}(A_4)$
& $\de_5^{\rm (TBA)}(D_4)$ & $\de_5^{\rm (TBA)}(E_6)$ \\ \hline
0.20 & --13.2856 & --12.7 &  &  &  &
       \\
0.25 & --6.49225 & --6.38 &        &  &   &
       \\
0.30 & --3.49933 & --3.45 & --3.55 & --3.57 &  &
       \\
0.35 & --2.03774 & --2.01 & --2.07 & --1.97 & --2.30 &
       \\
0.40 & --1.29349 & --1.281 & --1.31 & --1.31 & --1.01 &
     --2.5 \\
0.45 & --0.93614 & --0.929 & --0.95 & --0.95 & --0.94  &
     --2.0 \\
0.50 & --0.82954 & --0.823 & --0.84 & --0.84 & --0.84 &
      --0.87 \\ \hline
\end{tabular}
}
\caption{$\de_5^{\rm (RA)}$ vs.
$\de_5^{\rm (TBA)}$ for $A_2,A_3, A_4, D_4$, and $E_6$ ATFTs.}
\end{table}

The agreement is relatively poor for $\de_5$ for the cases with 
high rank such as $E_6$ partly due to the numerical errors 
in higher order calculations. 
Another reason comes from the fact that neglected terms in the 
$1/l$ expansion (the order of ${\cal O}(1/l^8)$ or higher) in 
Eq.(\ref{ceff}) may not be sufficiently small 
compared with terms with $\de_5$. 
However, one can in principle reduce these errors by
increasing the accuracy of the numerical calculations.
There are also corrections of ${\cal O}(R^{\gamma})$ to the expansion
of $c_{\rm eff}(R)$ in power series of $1/l$ which increase
as $B$ goes to zero. 
This explains why the discrepancies in the tables increase as $B$ decreases.

In Fig.2, we also plot the scaling functions $c_{\rm eff}(R)$ as a 
function of $R$ setting ${\overline m}=1/2$ for different ATFTs; 
first, using numerical solutions of the TBA equations 
and, second, using Eqs.(\ref{quantization}) and (\ref{grenergy}) based 
on the reflection amplitudes.
To compare the same objects, we added to the second case the contribution 
from the bulk vacuum energy term \cite{DDV,fateev} 
\beq
\Delta c_{\rm eff}={3{\overline m}^2R^2\over{2\pi}}
{\sin(\pi/h)\over{\sin(\pi B/h)\sin(\pi(1-B)/h)}}
\eeq
which becomes significant for $R\ge 0.01$.
For each model, the two curves are almost identical without any
noticeable difference in the graphs even upto $R\sim{\cal O}(1)$.

\begin{figure}
\rotatebox{-90}{\resizebox{!}{15cm}{\scalebox{0.1}{%
{\includegraphics[4cm,2cm][22cm,25cm]{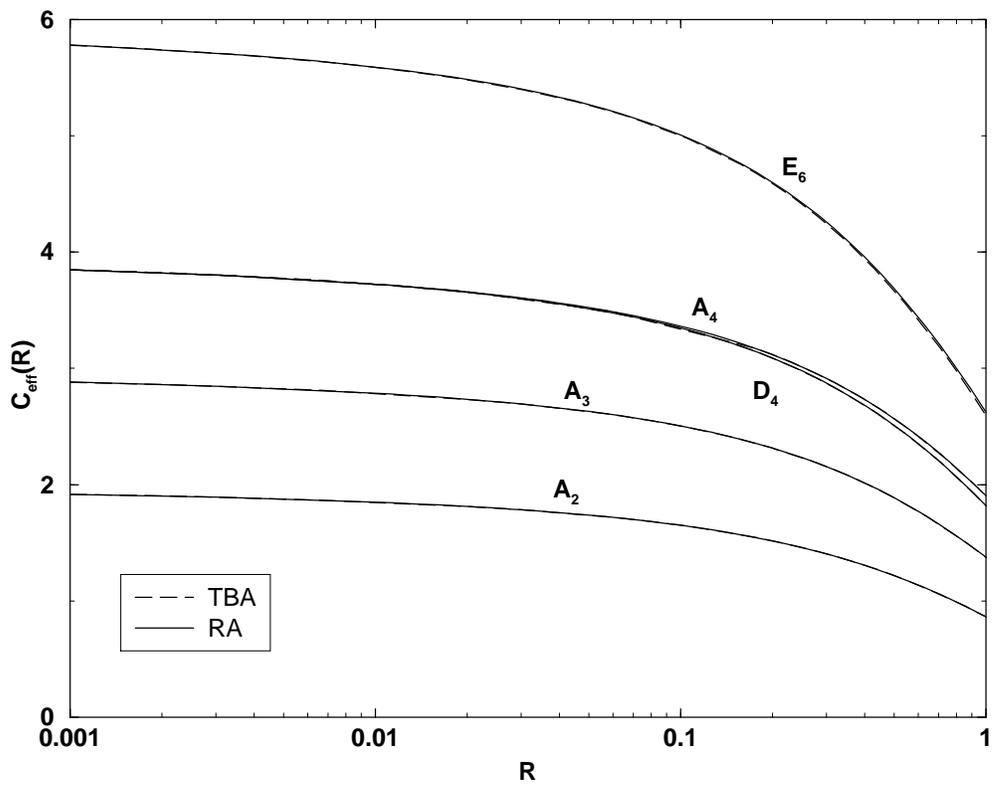}}}}}
\caption{Plot of $c_{\rm eff}$ for $A_2, A_3, A_4, D_4, A_6$,  
and $E_6$ ATFTs at $B=0.5$.} 
\end{figure}

\newpage

\section{Concluding remarks}

In the main part of this paper we considered the UV asymptotics
of the ground state energy in ATFTs. The main objects  which were used
for this analysis were the reflection amplitudes (\ref{rhats}) of NATTs.
As it was mentioned in sect.\ 2 these objects also play a crucial
role in the calculation of the one poin functions in perturbed CFT.
The one point functions of the exponential fields

\[
\G(\bds{a})= \langle \exp \bds{a}\cdot \bds{\varphi} \rangle
\]
in ATFTs can be reconstructed from the
reflection amplitudes (\ref{rhats}). It followes from the results of the 
paper \cite{FLZZ} that these functions satisfy to the functonal
equations similar to the relations (\ref{reflection}, \ref{reflection2}). 
For the ADE series of ATFTs these functions were calculated in \cite{fateev2}
and have the form:

\[
\G(\bds{a}) = \left(\frac{\bar m k(G)
\Gamma\left(\frac{1}{(1+b^2)h}\right)
\Gamma\left(1+\frac{b^2}{h(1+b^2)}\right)
}{2\Gamma(1/h)}\right)^{-a^2}
\exp{\T}(\bds{a})
\]
here
\[
\T(\bds{a})= \int\frac{dt}{t}\left[\psi(t)\,\sum_{\alpha>0}
\sinh(ba_{\bds{\alpha}} t)\,\sinh\left((b(a-2Q)_{\bds{\alpha}}+
h(1+b^2))t\right)+
e^{-2t} a^2\right]
\]
where  $k(G)$. $a_{\bds{\alpha}}$, $Q_{\bds{\alpha}}$ are defined by 
Eqs.\ (\ref{kofg}), (\ref{qalpha}) and
\[
\psi(t) = - \frac{\sinh((1+b^2)t)}{\sinh t \sinh(b^2t)\,\sinh((1+b^2)ht)}
\]

We suppose to discuss the application of these functions to
the analysis of ATFTs and related perturbed CFT in another
publication.

In the previous sections we studied the UV asymptotics for
the simply laced ATFTs. It looks interesting to extend this
consideration to the case of dual pairs of non simply lased
ATFTs. To do this it is necessary to generalize the quantization
condition (\ref{quantization}) and the relations (\ref{mmu}) between the 
parameters of the action and masses of the particles. The work in this 
direction is in progress.

\newpage
\section*{Appendix: $A_{n-1}$ Quantum Toda Chain}

In this Appendix we obtain the quantization conditions for 
one-dimensional $A_{n-1}$ quantum Toda chain using the approach
based on the integrability of this theory.
The quantum Toda chain corresponds to the semiclassical ($b\to 0$)
approach to the zero-mode dynamics of the ATFT, however, the
structure of quantization condition for this system is similar
to Eq.(\ref{quantization}).

The quantum Toda chain is described by the Hamiltonian
\beq
H=\half\sum_{i=1}^{n}p_i^2+\kap^2\sum_{i=1}^{n}e^{b(q_i-q_{i+1})}
\label{todachain}
\eeq
where $q_{n+1}\equiv q_1$.
The model (\ref{todachain}) is integrable. 
The diagonalization of the Hamiltonian was done in a remarkable
paper by Sklyanin \cite{Skl} who used the quantum inverse
scattering method to obtain the Bethe ansatz (BA) equations for the 
eigenvalues of $H$. 
Here we show that for $\kap<<1$ these BA equations can be represented
in the form similar to Eq.(\ref{quantization}).

The Lax operator for Toda chain (\ref{todachain}) has the form
\beq
L_i(u)=\left(
\begin{array}{cc}
u-p_i&-\kap e^{bq_i}\\
\kap e^{-bq_i}&0
\end{array}
\right).
\label{lax}
\eeq
The commuting family of the transfer matrices $T(u)$ corresponding
to the Lax operator Eq.(\ref{lax}) can be written as:
\beq
T(u)={\rm Tr}\left(L_1(u)\ldots L_n(u)\right),\qquad
\left[T(u),T(u')\right]=0.
\eeq
The operators $T(u)$ is a polynomial of degree $n$ with the 
coefficients:
\beq
T(u)=u^{n}+T_1u^{n-1}+T_2u^{n-2}+\ldots
\eeq
where $T_1=-\sum_{i=1}^{n}p_i=P$ and $T_2=-H+P^2/2$.
The BA equations for the eigenvalues $t(u)$ of the operator
$T(u)$ can be formulated in terms of eigenvalues $Q(u)$
of Baxter's $Q$-operator \cite{Baxt}.
Namely, these equations have the form \cite{Skl}
\beq
t(u)Q(u)=\kap^{n}\left(i^{n}Q(u+ib)+i^{-n}Q(u-ib)\right)
\label{baxterQ}
\eeq
where $t(u)$ is a polynomial of $u$, which can be written as
\beq
t(u)=\prod_{i=1}^{n}(u-v_i),\quad{\rm with}\quad
P=\sum_{i=1}^{n}v_i,\quad E=\sum_{i=1}^{n}v_i^2/2,
\eeq
and $Q(u)$ is an entire function of $u$.

With these two conditions Eq.(\ref{baxterQ}) is rather complicated
to be solved analytically. 
Even for $n=2$ it can be only reduced to the Mathieu equation.
Here we consider the approximate solution of Eq.(\ref{baxterQ})
with accuracy ${\cal O}(\kap^n)$. 
Namely, we impose the condition that function $Q(u)$ is a meromorphic
function with the absolute values of residues at the poles
${\cal O}(\kap^n)$.

The function $Q(u)$ with this analyticity property which satisfies
to Eq.(\ref{baxterQ}) with accuracy ${\cal O}(\kap^n)$ can be
written in the form
\beq
Q(u)=\exp\left({iun\log\kap\over{b}}\right)
\prod_{i=1}^{n}\Ga\left({i(v_i-u)\over{b}}\right)+
\exp\left(-{iun\log\kap\over{b}}\right)
\prod_{i=1}^{n}\Ga\left({i(u-v_i)\over{b}}\right).
\eeq
The function $Q(u)$ has the poles at real $u=v_i$ with the residues
${\cal O}(1)$ and the poles in the complex plane with the residues
${\cal O}(\kap^n)$.
To satisfy the analyticity condition we should cancel the poles
at $u=v_j$.
In this way we arrive at the following equations for parameters $v_j$:
\beq
\exp\left({2inv_j\log\kap\over{b}}\right)
\prod_{i\neq j}{\Ga\left(1+{i(v_i-v_j)\over{b}}\right)\over{
\Ga\left(1-{i(v_i-v_j)\over{b}}\right)}}=(-1)^{n-1}
\eeq
or taking the logarithm on both sides:
\beq
-{2nv_j\log\kap\over{b}}={1\over{i}}\sum_{i\neq j}
\log{\Ga\left(1+{i(v_i-v_j)\over{b}}\right)\over{
\Ga\left(1-{i(v_i-v_j)\over{b}}\right)}}+\pi I_j
\label{laxbae}
\eeq
where $I_j$ is the set of different integers, odd (even) for
even (odd) $n$.

The minimal set of numbers $I_j$ corresponding to the ground state 
with center of mass momentum $P=0$ can be written as:
\beq
I_j=2j-n-1,\quad j=1,\ldots,n.
\eeq
One can easily see that with this set of numbers $I_j$ 
Eq.(\ref{laxbae}) is similar to Eq.(\ref{quantization}) which
has additional term $\De\varphi(v_j)$ in the RHS:
\[
\De\varphi(v_j)=\sum_{i\neq j}{1\over{i}}
\log{\Ga\left(1+i(v_i-v_j)b\right)\over{\Ga\left(1-i(v_i-v_j)b\right)}}
\]
coming from the quantum renormalization of reflection amplitudes 
in the two-dimensional NATT.

\section*{\bf Acknowledgement}

We thank F. Smirnov and Al. Zamolodchikov for valuable discussions. 
CA thanks Freie Univ. in Berlin and Univ. Montpellier II for hospitality.
This work is supported in part by Alexander
von Humboldt foundation and MOST 98-N6-01-01-A-05 (CA), 
Korea Research Foundation 1998-015-D00071 (CR), and
KOSEF 1999-2-112-001-5(CA,CR). 
VF's work is supported in part by the EU under contract ERBFMRX CT960012.

\end{document}